\documentclass{aastex}
\usepackage{emulateapj5}

\newcommand{\BP}{Ballesteros-Paredes}

\newcommand{\Mcol}{M_{\rm col}}

\newcommand{\tcl}{\tau_{\rm cl}}

\newcommand{\VS}{V\'azquez-Semadeni}

\slugcomment{Submitted to ApJL}

\shorttitle{Star Formation Efficiency in Driven, Supercritical Clouds}
\shortauthors{\VS, Kim \& \BP}

\begin{document}


\title{Star Formation Efficiency in Driven, Supercritical, Turbulent Clouds}



\author{Enrique \VS\altaffilmark{1}, Jongsoo Kim\altaffilmark{2} and
Javier \BP\altaffilmark{1}}
\altaffiltext{1}{Centro de Radioastronom\'ia y Astrof\'isica (CRyA), UNAM,
Apdo. Postal 72-3 (Xangari), Morelia, Michoac\'an 58089, M\'exico
(e.vazquez, j.ballesteros@astrosmo.unam.mx)}
\altaffiltext{2}{Korea Astronomy and Space Science Institute (KASI), 61-1,
Hwaam-Dong, Yusong-Ku, Daejeon, 305-348, Korea
(jskim@kasi.re.kr)}


\begin{abstract}

We present measurements of the star formation efficiency (SFE) in 3D
numerical simulations of driven turbulence in supercritical, ideal-MHD,
and non-magnetic regimes, characterized by their mean normalized
mass-to-flux ratio $\mu$, all with 64 Jeans masses and similar rms Mach
numbers ($\sim 10$). In most cases, the moderately supercritical runs
with $\mu = 2.8$ have significantly lower SFEs than the non-magnetic
cases, being comparable to observational estimates for whole molecular
clouds ($\lesssim$ 5\% over 4 Myr). Also, as the mean field is
increased, the number of collapsed objects decreases, and the median
mass of the collapsed objects increases.  However, the largest
collapsed-object masses systematically occur in the weak-field case $\mu
= 8.8$. The high-density tails of the density histograms in the
simulations are depressed as the mean magnetic field strength is
increased. This suggests that the smaller numbers and larger masses of
the collapsed objects in the magnetic cases may be due to a greater
scarcity and lower mean densities (implying larger Jeans masses) of the
collapse candidates. In this scenario, the effect of a weak field is to
reduce the probability of a core reaching its thermal Jeans mass, even
if it is supercritical. We thus suggest that the SFE may be
monotonically reduced as the field strength increases from zero to
subcritical values, rather than there being a discontinuous transition
between the sub- and supercritical regimes, and that a crucial question
to address is whether the turbulence in molecular clouds is driven or
decaying, with current observational and theoretical evidence favoring
(albeit inconclusively) the driven regime.

\end{abstract}


\section{Introduction} \label{sec:intro}


The fraction of a molecular gas mass that is converted into stars, known
as the star formation efficiency (SFE), is known to be small,
ranging from a few percent for entire molecular cloud complexes
(e.g., Myers et al.\ 1986) to 10--30\% for cluster-forming cores
(e.g., Lada \& Lada 2003), even though molecular clouds in general have
masses much larger than their thermal Jeans masses, and should therefore
be undergoing generalized gravitational collapse if no other processes
prevented it \citep{ZP74}. Thus, this reduction
of the mass that is deposited in collapsed objects needs to be accounted
for by models of the star formation process. 

In the so-called ``turbulent'' model of
star formation (see, e.g., the reviews by \VS\ et al.\ 2000; Mac Low \&
Klessen 2004; \VS\ 2004), the low efficiency arises because the
supersonic turbulence within the clouds, while contributing to global
support, generates large-amplitude density
fluctuations (clumps and cores), some of which may themselves become locally
gravitationally unstable and collapse in times much shorter than the
cloud's global free-fall time. Thus, collapse occurs locally rather
than globally, and involves only a fraction of the cloud's total mass. 
This fraction constitutes the SFE, and depends on the global properties
of the cloud, such as its rms Mach number, the number of Jeans masses it
contains, and the turbulence driving scale 
(L\'eorat, Passot \& Pouquet 1990; Klessen, Heitsch \& Mac Low 2000),
or, perhaps more physically, on the scale at which the turbulent
velocity dispersion becomes subsonic \citep{Pad95,VBK03}. Nevertheless,
the SFE appears to still be too large in non-magnetic configurations,
being $\sim 30$\% in simulations with realistic parameters
\citep[e.g.,][]{KHM00,VBK03}, 
and it is important to investigate the contribution of 
the magnetic field in further reducing the SFE. 
This remains an open issue.
In three-dimensional (3D) simulations
of driven, self-gravitating, ideal MHD turbulence, \citet{HMK01} studied the
evolution of the mass fraction in collapsed objects (a measure of the
SFE) in supercritical and 
non-magnetic cases. They reported, however, that any systematic trends
with the field strength that might have been present in their
simulations were blurred by statistical fluctuations from one realization to
another. 

More recently, \citet[][ hereafter LN04]{LN04} and \citet[][ hereafter
NL05]{NL05}, have measured the SFE in two-dimensional (2D) simulations of 
\emph{decaying} turbulence including ambipolar diffusion (AD) and (in
NL05) a model prescription for outflows. LN04
found that the initial turbulence can accelerate the formation and
collapse of cores within the clouds. NL05 concluded that
SFEs comparable to those of whole molecular clouds (a few percent)
required moderately 
subcritical conditions, while moderately supercritical cases gave
efficiencies comparable to cluster-forming cores ($\sim 20$\%). However,
since their simulations were done in a decaying regime and in a closed
numerical box, the SFEs they measured are probably upper limits.
Real clouds may not be in a decaying regime (\VS\ et al.\ 2005,
hereafter Paper I; see also
the discussion in \S \ref{sec:disc}) and 
moreover probably undergo partial dispersion in
response to their turbulent energy contents, as in the simulations of
\citet{CB04} and \citet{CBZB05}, thus reducing the amount of mass
available for collapse. 

Paper I studied the
formation, evolution and collapse of the cores formed in 3D simulations
of \emph{driven} MHD turbulence,
albeit neglecting AD. Due to this setup, no collapse could occur in
subcritical cases. Although no quantitative measurements
of the SFE were reported there, 
a trend towards decreasing collapse rates with increasing field
strengths was clearly observed in supercritical and non-magnetic
cases. This was \emph{not} due to longer individual 
core collapse times in the magnetic cases with respect to the
non-magnetic one, since the timescales for formation and collapse of
the cores were similar in both cases. Instead, the
reduction of the SFE was apparently due to a
reduced formation rate of collapsing objects in the magnetic
simulations in comparison with the non-magnetic case. 

The goal of the
present paper is to report quantitative measurements of the SFE and of
the numbers and masses of collapsed objects forming in the
simulations of Paper I and three new sets of similar ones, as a function
of the mean field strength. In order
to overcome the difficulties encountered by \citet{HMK01}, we study these
variables at fixed settings of the random turbulence driver. 

\section{Numerical method, simulations and procedure} \label{sec:meth_simul}

We refer the reader to Paper I for details on the simulations and
resolution considerations. 
%
%
We consider four sets of simulations at a resolution of $256^3$ keeping
all physical parameters constant, except for the mass-to-magnetic flux
ratio $\mu$, which we vary to investigate the effects of the magnetic
field. One set consists of the four simulations presented in Paper I
with rms Mach number $M\approx 10$, Jeans number $J\equiv L/L_{\rm J} =
4$ (where $L$ is 
the numerical box size and $L_{\rm J}$ is the Jeans length), and $\mu=$
0.9, 2.8, 8.8 and $\infty$, corresponding to subcritical, moderately
supercritical, strongly supercritical and non-magnetic cases. We refer
to this set by the label ``Paper I''. 
We also consider three more sets
of three similar simulations each (with $\mu = 2.8$, 8.8 and $\infty$), but varying the seed of the random turbulence
driver. We label the sets by their seed numbers, as ``Seed = 0.1'',
``Seed = 0.2'', and ``Seed = 0.3''. The driving is
computed in Fourier space, and applied at the largest scales in the
simulation ($\sim 1/2$ of the box length), so that it
is not expected to be the main driver of the local evolution of the
clumps and cores, because, on the scales of the cores, the applied force
is nearly uniform, and its main effect should just be to push the cores
around without severely distorting them.

Although the simulations are scale-free, for reference, a convenient set
of physical units is $n_0  
= 500$ cm$^{-3}$, $u_0 = c_{\rm s} = 0.2$ km s$^{-1}$, $L_0 = L = 4$ pc,
and $t_0 = L_0/u_0 = 20$ Myr. The latter is the sound crossing time
across the box. Taking the mean molecular mass as $m = 2.4 m_{\rm H}$,
the numerical box then contains $1.86 \times 10^3 $M$_\sun$. The mean
field strengths for the $\mu = 0.9$, 2.8, 8.8 and $\infty$ cases are 
respectively $B_0 = 45.8$, 14.5, 4.58, and 0 $\mu$G, corresponding to values of
$\beta$, the ratio of thermal to 
magnetic pressure, of $\beta = 0.01$, 0.1, 1 and $\infty$. The simulations are
run for 0.5 code time units (10 Myr) before turning on the self-gravity.

As a measure of the SFE, we consider the evolution of the collapsed mass
fraction $\Mcol$ of the simulations. By this we
mean gas that is at densities $n > 500 n_0$, since in Paper I we noticed
that once an object reached densities $\sim 300 
n_0$ it was always already on its way to collapse, a fact which is
confirmed by the fact that these objects never disperse during the
subsequent evolution of the simulations. Throughout the paper, we refer
exclusively to collapsed objects rather than stars, because our spatial
resolution 
is clearly insufficient to determine whether a collapsing object
eventually breaks up into more fragments to form several stars. This is
likely to be the case for the most massive collapsed objects, with
masses up to $\sim 100$ M$_\sun$. Thus, the masses reported in fig.\
\ref{fig:core_data} should not be necessarily interpreted as individual
stellar masses, and may well be cluster masses. Also note that in runs
that form a single collapsed object, the plots of $\Mcol$ vs.\ time
really represent the accretion history onto that object, rather than the
continuous formation of new collapsed objects. 


As a representative cloud lifetime we take $\tcl = 4$ Myr, a time scale that
agrees with the estimate of \citet{BHRB04} of 3--5 Myr and also with the
two-turbulent-crossing-times criterion used by \citet{VBK03}. Quoted
values of the SFE refer to the collapsed mass fraction at $t=\tcl$. Note,
however, that, because of numerical problems when the density
contrast becomes too large, not
all simulations reach this time, although this will not be a limitation for
the conclusions we will draw.

\section{Results} \label{sec:results}

\begin{figure}
\epsscale{.75}
\plotone{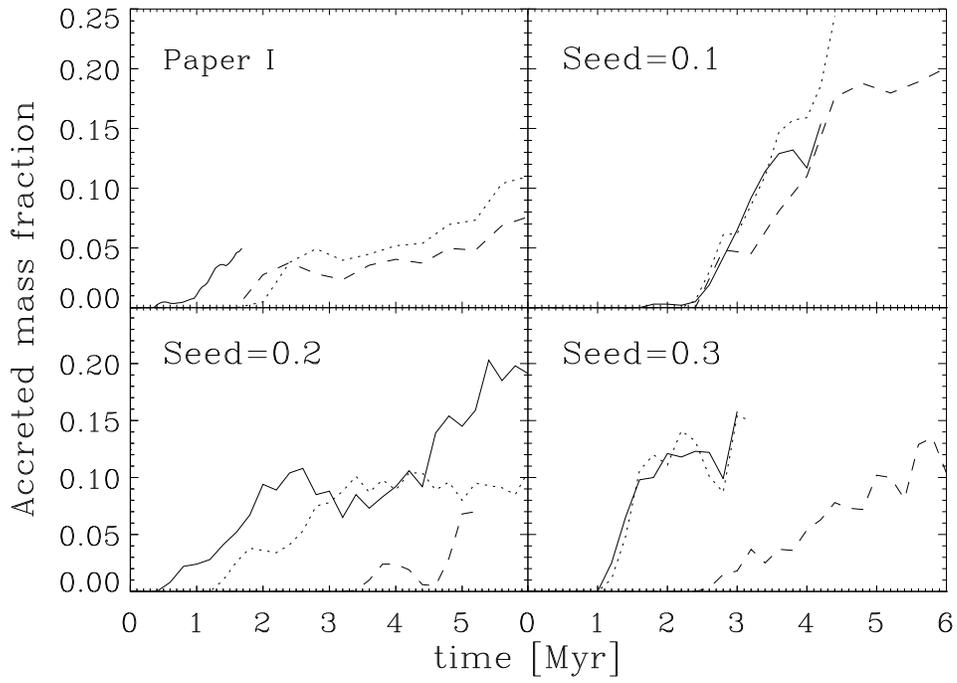}
\caption{Evolution of the accreted mass fraction for the four sets of runs
considered. {\it Solid} lines denote $\mu
= \infty$, {\it dotted} lines denote $\mu
=8.8$, and {\it dashed lines} denote $\mu =2.8$.} 
\label{fig:macc_vs_t}
\end{figure}

\begin{figure}
\epsscale{.75}
\plotone{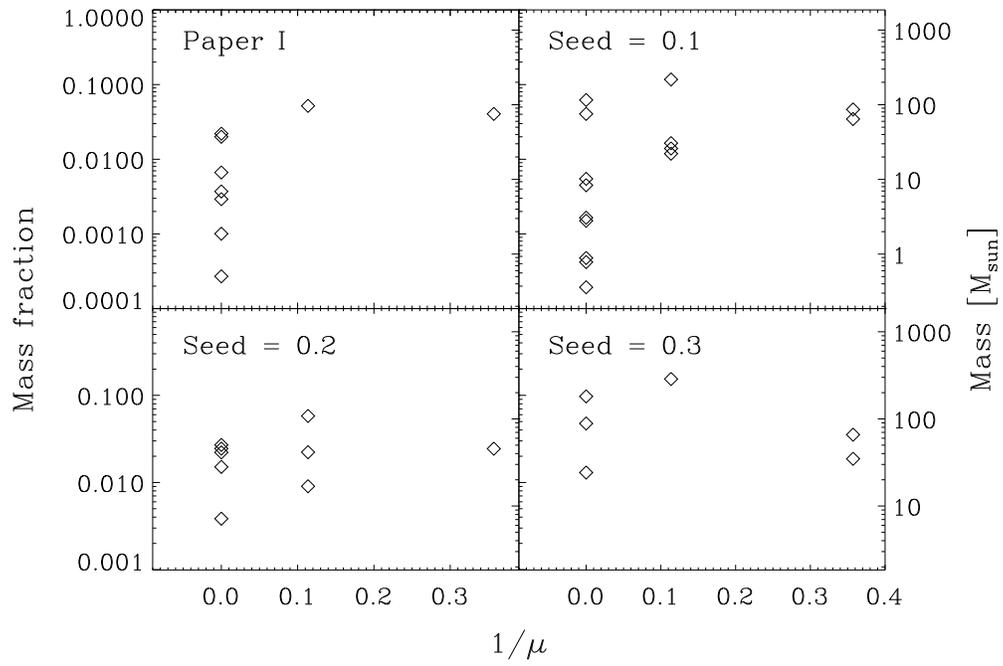}
\caption{Masses of the collapsed objects (objects with densities $n >
500 n_0$) versus the inverse of the simulation mass-to-flux ratio for
the four sets of runs. The left vertical axis gives the masses as
fractions of the total mass in the simulation, and the right axis gives
them in solar masses, according to the adopted normalization.}
\label{fig:core_data}
\end{figure}

Figure \ref{fig:macc_vs_t} shows the evolution of the accreted mass
fraction for the four sets of runs we consider. 
The moderately supercritical runs, with $\mu = 2.8$, have generally
lower SFEs than both the strongly supercritical ($\mu = 8.8$) and
non-magnetic ($\mu = \infty$) runs. In turn, the $\mu=8.8$ cases have
SFEs that are generally very similar to those of the non-magnetic
runs, except in the runs from Paper I. 
It is also noteworthy that the $\mu = 2.8$ cases have SFEs $\approx 0.04$,
0.12, 0.025 and 0.05, respectively for each set. Thus, in three out of the four
statistical realizations, $\Mcol \leq 5$\% at $t=\tcl$, in reasonable
agreement with the observed SFEs at the level of global molecular
clouds. 

What is even more interesting is the different way in which the
magnetic and the non-magnetic runs reach their respective collapsed
fractions. Figures 5 and 9 in Paper I, and their corresponding
animations in the electronic version, show
that, while the magnetic runs do so with one or two relatively massive
collapsed objects, the non-magnetic run does so with several objects,
many of them with low masses. The same trend is observed in the three
additional sets of 
statistical realizations considered in the present paper. In fig.\
\ref{fig:core_data} we show the masses of the
individual collapsed objects ($n > 500 n_0$) for all the runs we
consider at $t=4$ Myr, except for those cases in which the simulation
terminated prematurely, in which case we plot the collapsed object
masses at the last timestep of the simulation. We see that the
non-magnetic runs typically produce many more collapsed objects than the
magnetic runs, and with mass distributions that extend to significantly lower
values. On half the cases, the minimum 
masses of the collapsed objects increase monotonically with
increasing field strength.

\begin{figure}
\epsscale{.75}
\plotone{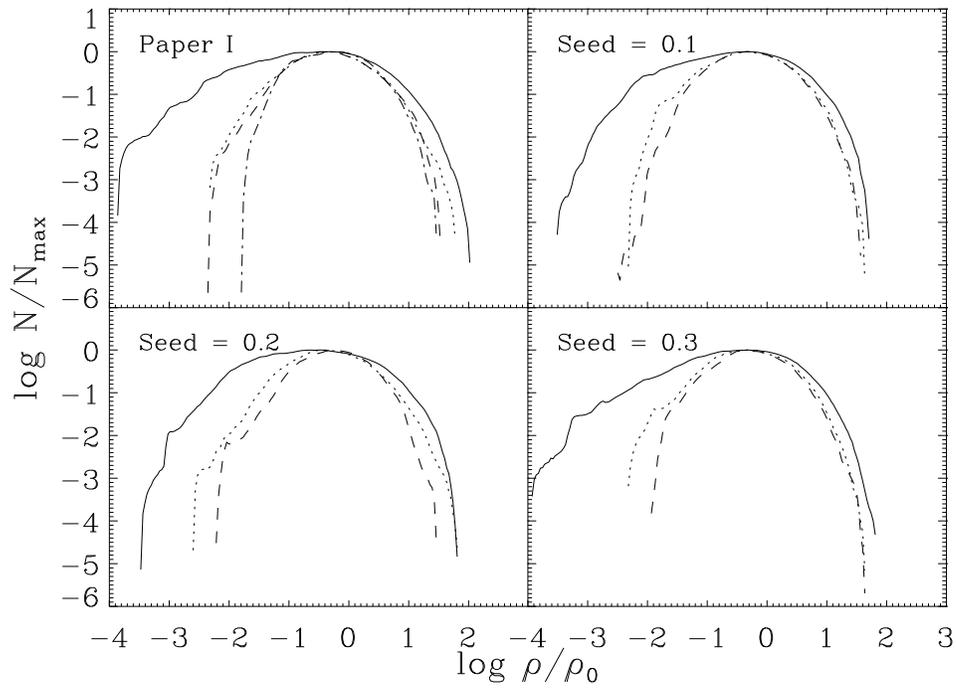}
\caption{Normalized density histograms for all the simulations
immediately before gravity is turned on. The line coding is as in fig.\
\ref{fig:macc_vs_t}. The {\it dot-dashed} line in the
``Paper I'' panel denotes the subcritical case with $\mu = 0.9$. 
} 
\label{fig:hists}
\end{figure}

One of the main results of Paper I was that the core formation+collapse time
scale was not significantly different between the magnetic and
non-magnetic cases. Thus, the above result on the masses and numbers of
the collapsed objects suggest that the decreased efficiency of the
magnetic cases in comparison to the non-magnetic ones arises from a
decreased probability of collapse events as the field strength
increases, rather than from an increase of the collapsing object
lifetimes. This suggestion is supported by the probability distribution
of the density fluctuations for the various runs. Figure \ref{fig:hists}
shows the histograms of the density values in each of the simulations,
averaged over the last five snapshots before turning on gravity. 
In this figure, the panel for
the runs from Paper I includes the subcritical run with $\mu = 0.9$, in
order to see the effect of the magnetic field in this case as well, even
though this run did not undergo collapse. 

Figure \ref{fig:hists} shows a clear trend towards decreasing
width and lower high-density tails with increasing mean field
strength (increasing $\mu$), at least over the range of magnetic field
strengths we have considered here. That is, the probability of producing
large density enhancements decreases with increasing magnetic field
strength. This is in agreement with previous results by \citet{PVP95},
\citet{OGS99}, \citet{HMK01} and \citet{BPML02}. Note, however, that in those
works a reverse trend towards increasing fluctuation amplitude was
observed at stronger values of the field, which we do not observe here,
again at least over the field strength range spanned by our simulations.

The trend towards lower and less extended high-density tails in the
histograms at larger field strengths is consistent with the trend of the SFE
to decrease and of the minimum collapsed masses to increase in the same
limit, since fewer density fluctuations can reach the threshold 
for collapse, and simultaneously, the local values of the Jeans mass are
larger, so that, in order to collapse, an object needs to acquire more
mass. In this picture, the field's effect on the SFE is only indirect,
through its 
modification of the density histogram, rather than by directly
increasing the minimum mass for collapse (i.e., by causing the
magnetically critical mass of the cores to be larger than their Jeans
masses). Indeed, in the cases studied in Paper I, examples of both
collapsing and non-collapsing cores were supercritical, and the
occurrence of collapse depended on whether they acquired the Jeans
mass. Similarly, \citet{Li_etal04} found that all the cores in their
supercritical simulations were supercritical.


Finally, it is worth noting that the most massive objects systematically
arise in the strongly supercritical cases ($\mu = 8.8$) in all four
simulation sets. 
The origin of this effect, as well as a test of the mechanism suggested
above for reducing the SFE and increasing the minimum masses will
require detailed 
measurements of the field morphology and the evolution of the energy balance
in the cores prior and during the onset of collapse, to be presented
elsewhere.

\section{Discussion} \label{sec:disc}

The above results can be placed in the context of previous studies. We
have found that the presence of a magnetic field can further
reduce the SFE with respect to the non-magnetic case, even in
supercritical configurations. 
Previous successful determinations of the effect of the magnetic field
have been restricted to decaying, 2D simulations (LN04; NL05). These
authors found that values of the SFE comparable to those observed in entire
molecular clouds \citep[a few percent;][]{Myers_etal86} required subcritical
environments and AD-mediated collapse, while supercritical environments
gave values of the SFE closer to those of cluster-forming cores
($\sim$ 15\% after 1 global free-fall time). 

Instead, in our simulations we have
found that a moderately supercritical environment and
reasonably realistic values of the Mach number (10) and of the Jeans
number (64 Jeans masses in a 4-pc cube) already give SFEs $\lesssim$5\%
in 3/4 of the cases studied after 0.8 global free-fall times (4 Myr), in
spite of having a mass-to-flux ratio more than twice larger than theirs. The
difference is probably due mainly to the choice of global setup (2D,
decaying, versus 3D, driven), since
a continuously-driven simulation maintains the turbulent support
throughout its evolution, while a decaying one loses it over time. 

In the simulations of LN04 and NL05, the main role of turbulence is to
accelerate the initial formation of the cores and other density structures,
which occurs on the turbulent crossing time rather than in the long AD
timescale. However, as the turbulence decays, its role in providing
support and producing further fragmentation (which induces \emph{local}
collapse involving small fractions of the total mass) progressively
decreases. Indeed, the turbulent Mach numbers in 
the simulations of NL05 had already decayed to values $\sim$ 2--3 by the
times the collapsed objects were forming. Thus, in their supercritical
cases, the entire bulk 
mass of the simulation is in principle available for collapse at long
enough times, although the residual turbulent fragmentation still drives
local collapse events first.

This suggests that a fundamental question in understanding the SFE in
molecular clouds is whether real molecular clouds are driven or 
decaying, and sub- or supercritical. Concerning the former dichotomy,
observational evidence tends to suggest that molecular clouds are
driven, as we discussed at length in Paper I; a few additional
considerations are as follows. If the turbulence is generated by
instabilities in the 
compressed layers as the clouds are forming \citep[e.g., ][]{Hunter_etal86,
Vishniac94, WF00}, the injection
of turbulent energy is likely to last for as long as the accumulation process
lasts. Afterwards, the cloud is likely to disperse, as indicated by
the facts that the gas has disappeared from star-forming regions after a
few Myr \citep{HBB01}, and that all clouds of comparable masses tend to have
comparable levels of turbulence \citep{HB04}. If the turbulence in
clouds were decaying, one would expect that clouds of a given mass
would exhibit a large scatter in their turbulence levels, contrary to
what is observed. The ``universal'' behavior of the turbulent level in
clouds reported by \citet{HB04} also suggests that the clouds 
are part of the global Galactic turbulent cascade, in which the key
process is a {\it statistically stationary} energy transfer among
scales, analogous to the classical Kolmogorv cascade. Thus,
driven-turbulence simulations may be a somewhat better approximation to
real clouds than decaying ones, although the standard Fourier-driving 
scheme is likely not to be the best model of the true injection
mechanism. Simulations with more realistic driving schemes are clearly
needed.

Concerning the subcritical vs.\ supercritical dichotomy, theoretical
arguments suggest that as a cloud is becoming predominantly molecular
and self-gravitating, it is also becoming supercritical
\citep{McK89,HBB01}. Recent observational evidence remains
inconclusive, and tends to suggest that both kinds are realized,
with probably some preponderance of supercritical ones \citep{Crut99,
Bourke01, Crut04}, although the uncertainties are large. The safest
assumption at this point appears to be that both regimes are realized
in molecular clouds.

Thus, the global picture that emerges is that a \emph{distribution} of
magnetic field strengths exists in the ensemble of molecular clouds
\citep{BV97}, and that the SFE decreases monotonically as the mean
field strength in the clouds is larger. This appears to be a
\emph{continuous} trend, rather than a sharp dichotomy between sub-
and supercritical regimes, as was the case in
the standard model of star formation \citep{SAL87, Mousch91}. 
The role of AD would then mostly be to just allow subcritical clouds to
participate in the star formation process, with the lowest SFEs of
the spectrum. A confirmation of this picture will require a systematic
study of the SFE in 3D, driven simulations including AD, to be presented
elsewhere.

Our second result is that the number of collapsed objects appears to
decrease, and their median mass appears to increase with increasing mean
field strength. This result is consistent with the observation that 
in the non-magnetic case a single clump appears to form
several collapsed objects, while in the magnetic cases a clump appears
to form a single object, as can be seen in the animations presented
in Paper I. This observation goes in line with the findings of
\citet{HMK01}, who noticed that non-magnetic simulations with
large-scale driving tended to form \emph{clusters} of collapsed objects,
while magnetic cases tended to form the collapsed objects in
a more scattered fashion. If collapsing objects in the
magnetic case are more massive, then clumps of a given mass can form
fewer objects. 
Detailed analysis of the evolution of individual cores will be
necessary to test these possibilities.

\acknowledgments
This work has received partial financial support from grants CONACYT
36571-E to E.V.-S. J.K.\ was supported by ARCSEC, of the
Korea Science and Engineering Foundation, through the SRC program. The
numerical simulations have been performed  
on the linux clusters at KASI (funded by KASI and ARCSEC) and at CRyA
(funded by the above CONACYT grant). 


\clearpage

\clearpage

\clearpage

\end{document}